\documentclass[conference,10pts,compsocconf]{IEEEtran}

\usepackage{graphicx,psfrag}
\usepackage{mathrsfs}
\usepackage{amsmath,amsfonts,amssymb}
\usepackage{multirow}
\usepackage{comment}
\usepackage{enumerate}
\usepackage{mathrsfs}
\usepackage{pifont}
\usepackage[normalem]{ulem}
\usepackage[utf8]{inputenc}

\def\NRtidal{\texttt{NRTidal}}
\usepackage{color}

\begin{document}

\title{High-resolution numerical relativity simulations of spinning binary neutron star mergers}

\author{
  \IEEEauthorblockN{%
    Tim Dietrich\IEEEauthorrefmark{1}\,\IEEEauthorrefmark{2},
    Sebastiano Bernuzzi\IEEEauthorrefmark{3}\,\IEEEauthorrefmark{4},
    Bernd Br\"ugmann\IEEEauthorrefmark{5},
    Wolfgang Tichy\IEEEauthorrefmark{6}
  }
  \IEEEauthorblockA{%
    \IEEEauthorrefmark{1}Max Planck Institute for Gravitational Physics (Albert Einstein Institute), Am M\"uhlenberg 1, Potsdam-Golm, 14476, Germany \\
    \IEEEauthorrefmark{2}Nikhef, Science Park 105, 1098 XG Amsterdam, The Netherlands\\  
    \IEEEauthorrefmark{3}Dipartimento di Scienze Matematiche Fisiche ed Informatiche, Universit\'a di Parma, I-43124 Parma, Italia \\
    \IEEEauthorrefmark{4}Istituto Nazionale di Fisica Nucleare, Sezione Milano Bicocca, gruppo collegato di Parma, I-43124 Parma, Italia \\
    \IEEEauthorrefmark{5}Theoretical Physics Institute, University of Jena, D-07743 Jena, Germany \\
    \IEEEauthorrefmark{6}Department of Physics, Florida Atlantic University, Boca Raton, FL 33431 US \\    
  }
}%end author

\IEEEoverridecommandlockouts
\IEEEpubid{\makebox[\columnwidth]{ 978-1-5386-4975-6 
\copyright2018 IEEE \hfill} \hspace{\columnsep}\makebox[\columnwidth]{ }}

\maketitle

\date{\today}

\begin{abstract}
The recent detection of gravitational waves and electromagnetic counterparts 
emitted during and after the collision of two neutron stars 
marks a breakthrough in the field of multi-messenger astronomy.
Numerical relativity simulations are the only tool to describe the
binary's merger dynamics in the regime when speeds are largest and
gravity is strongest. 

In this work we report state-of-the-art binary neutron star simulations  
for irrotational (non-spinning) and spinning configurations. The main
use of these simulations is to model the gravitational-wave signal. Key 
numerical requirements are the understanding of the convergence
properties of the numerical data and a detailed error budget.
The simulations have been performed on different HPC clusters, 
they use multiple grid resolutions, and are based on 
eccentricity reduced quasi-circular initial data. 
We obtain convergent
waveforms with phase errors of $0.5-1.5\text{rad}$ accumulated over $\sim 12$ orbits to merger. 
The waveforms have been used for the construction of a phenomenological waveform
model which has been applied for the analysis of the recent binary neutron star detection.
Additionally, we show that the data can also be used to 
test other state-of-the-art semi-analytical waveform models. 
\end{abstract}

\section{Introduction}
\label{sec:intro}

Neutron stars (NSs) are among the most compact objects in the universe
with central densities multiple times higher than nuclear density. 
Similar conditions are unreachable on earth which makes
NSs an exceptional laboratory to test nuclear physics predictions. 
In particular the merger of two NSs allows the study of the high 
density region of the equation of state (EOS) governing 
NS matter. 
In addition, NS mergers also allow us to 
reveal the central engine for luminous short Gamma ray bursts (sGRBs), 
to understand the origin of heavy elements in the universe, 
which after their creation produce the optical and near-infrared
EM counterparts, called kilonovae, and 
to test astrophysical predictions about binary populations. \\

The first detection of gravitational waves (GW) combined with 
an observation of a sGRB and a kilonova 
marks a breakthrough in the field of 
multi-messenger astronomy~\cite{TheLIGOScientific:2017qsa,GBM:2017lvd}. 
It is expected that with the increasing sensitivity
of advanced GW interferometers multiple GW detections 
of merging BNSs will be made in the next years~\cite{Abbott:2016ymx}. 
In order to extract information from the data, 
the measured signal is cross-correlated 
with a GW template family to obtain a ``best match''. 
This requires accurate GW templates to relate 
the source properties to the observed GW signal and consequently 
a detailed analysis of the compact binary coalescence 
close to the moment of merger. 

While an analytical approach to the two-body dynamics in general
relativity is possible for the stage in which the bodies 
are well-separated, a numerical solution of the 
field equations, dealing with all their nonlinearities, 
is needed for a faithful description of the last few orbits. However, 
general relativistic simulations are computationally challenging 
and expensive. The main reasons are: 
(i)~the nonlinearity of the equations, 
(ii)~the intrinsic multi-scale character of the problem 
(covering the neutron star interior and the radiation zone), 
(iii)~no symmetries can be
exploited for generic binary simulations 
(3D in space plus time),
(iv)~the appearance of shocks and discontinuities in the matter fields. 
Over the last years, significant progress has been made simulating BNSs, 
with detailed descriptions of physical processes
as finite temperature EOSs, magnetic fields, neutrino transport, e.g.~\cite{Dionysopoulou:2012zv,Sekiguchi:2015dma,Palenzuela:2015dqa,Kiuchi:2017zzg,Foucart:2017mbt},
with new numerical techniques such as discontinuous 
Galerkin methods~\cite{Bugner:2015gqa,Kidder:2016hev,Dumbser:2017okk} 
and high-order convergent schemes~\cite{Radice:2013hxh,Bernuzzi:2016pie}, 
and with the possibility 
to study a larger region of the BNS parameter space 
with spinning~\cite{Tichy:2012rp}, eccentric~\cite{Moldenhauer:2014yaa}, 
and high-mass ratio~\cite{Dietrich:2015pxa} configurations. \\

In this article we present recent state-of-the-art numerical relativity 
simulations of irrotational and spinning BNSs employing high-order convergent methods. 
Notice that although spin is one of the elementary
parameters of a binary system, most studies of BNSs in numerical 
relativity have not considered spinning NSs, 
see e.g.~\cite{Bernuzzi:2013rza,Tacik:2015tja,Dietrich:2016lyp} for 
a few exceptions of spinning BNSs described in a consistent approach.
However, it is important to include spin in simulations since 
NSs in binaries are spinning objects~\cite{Lorimer:2008se}.
The most famous example is the double pulsar PSR J0737-3039 for which
the orbital period and both spin periods are known~\cite{Burgay:2003jj,Lyne:2004cj}.
In addition to the novelty of describing spinning NSs, 
we make use of the fact that recently it became possible to produce NS binaries with 
arbitrary eccentricity, in particular, with low eccentricities. 
While standard techniques lead to configurations with eccentricities of $\sim0.01$, 
astrophysical BNS systems will have significantly smaller eccentricities. 
Not reducing the eccentricity leads to spurious oscillations in the phase evolution
which reduces the quality of the waveform and hampers waveform model development. \\

The article is structured as follows: 
We discuss the configurations and 
numerical methods in Sec.~\ref{sec:methods}, the code performance in 
Sec.~\ref{sec:performance}, and we assess 
the accuracy of the gravitational 
waveforms in Sec.~\ref{sec:accuracy}. 
We focus in particular on the GW phase 
which is the most important quantity 
for waveform modeling and find that 
to date the presented simulations
are the most accurate simulations of spinning BNSs, see 
e.g.~\cite{Bernuzzi:2011aq,Radice:2013hxh,Hotokezaka:2015xka,Kiuchi:2017pte} for high-resolution simulations 
of irrotational BNSs. We present applications of the waveforms in Sec.~\ref{sec:applications} and 
conclude in Sec.~\ref{sec:conclusion}.

\section{Numerical Methods \& Fiducial binaries}
\label{sec:methods}

\subsection{Binary configurations}
\label{sec:methods:config}

\renewcommand{\arraystretch}{1.2}
\begin{table*}[h]
  \centering    
  \caption{BNS configurations. 
    The first column defines the name of the configuration
    with the notation: EOS$_{M^A}^{\chi^A}$.
    The subsequent columns describe:
    the EOS~\cite{Read:2008iy}, 
    the NS' individual masses $M_{A,B}$, 
    the stars' dimensionless spins $\chi_{A,B}$,
    the effective dimensionless coupling constant 
    $\kappa^T_{\rm eff}$ of the binary,
    the residual eccentricity~\cite{Dietrich:2015pxa},  
    the resolutions employed for the evolution grid, 
    and the grid resolutions $h_6$ in the finest level, where 
    we set $G=c=M_\odot=1$. 
    }     
\begin{small}
\begin{tabular}{l||ccccc|cc}        
\hline
\hline
  Name & EOS & $M_{A,B}$ & $\chi_{A,B}$ & $\kappa^T_{\rm eff}$ & $e [10^{-3}]$ & $n_6$ & $h_6$  \\
     \hline
MS1b$_{1.35}^{-0.10}$ & MS1b & 1.3504 & -0.099 & 288.0 & 1.8 & (64,96,128,192) & (0.291,0.194,0.145,0.097) \\
MS1b$_{1.35}^{0.00}$  & MS1b & 1.3500 & +0.000 & 288.0 & 1.7 & (64,96,128,192) & (0.291,0.194,0.145,0.097)  \\
MS1b$_{1.35}^{0.10}$  & MS1b & 1.3504 & +0.099 & 288.0 & 1.9 & (64,96,128,192) & (0.291,0.194,0.145,0.097) \\
MS1b$_{1.35}^{0.15}$  & MS1b & 1.3509 & +0.149 & 288.0 & 1.8 & (64,96,128,192) & (0.291,0.194,0.145,0.097) \\
\hline
 H4$_{1.37}^{0.00}$   & H4   & 1.3717 & +0.000 & 190.0 & 0.9 & (64,96,128,192) & (0.249,0.166,0.125,0.083) \\
 H4$_{1.37}^{0.14}$   & H4   & 1.3726 & +0.141 & 190.0 & 0.4 & (64,96,128,192) & (0.249,0.166,0.125,0.083) \\
\hline
SLy$_{1.35}^{0.00}$   & SLy  & 1.3500 & +0.000 & 73.5  & 0.4 & (64,96,128,192,256) & (0.234,0.156,0.117,0.078,0.059)\\     
SLy$_{1.35}^{0.05}$   & SLy  & 1.3502 & +0.052 & 73.5  & 0.4 & (64,96,128,192) & (0.234,0.156,0.117,0.078)\\     
SLy$_{1.35}^{0.11}$   & SLy  & 1.3506 & +0.106 & 73.5  &0.7 & (64,96,128,192) & (0.234,0.156,0.117,0.078)\\ 
\hline      
\hline
     \end{tabular}
     \end{small}
 \label{tab:config}
\end{table*}

In total nine different physical configurations have been simulated 
employing three different EOSs (MS1b, H4, SLy), see~\cite{Read:2008iy} for more details 
about the used EOSs. 
The intrinsic rotation of the NSs is characterized by the dimensionless spin of the 
stars $\chi_{A,B} = S_{A,B}/M_{A,B}^2$ with $S_{A,B}$ being the angular momentum 
of the stars and $M_{A,B}$ their masses in isolation. 
All spins are either aligned $(\chi>0)$
or anti-aligned $(\chi<0)$ to the orbital angular momentum. 
We summarize the physical parameters of the 
simulated BNSs in Table~\ref{tab:config}. 

\begin{figure}[t]
   \includegraphics[width=0.48\textwidth]{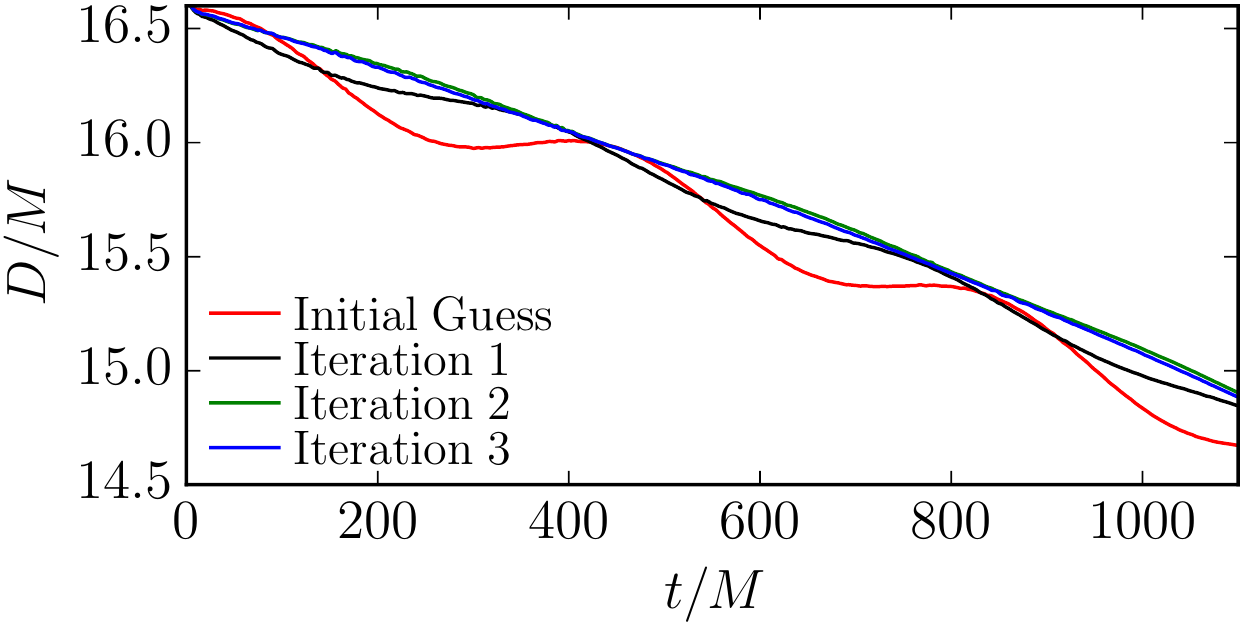}
   \caption{ \label{fig:ecc-red}
   Eccentricity reduction procedure for SLy$_{1.35}^{0.05}$.
   We present the proper distance of the system as a function of time. 
   The artificial eccentricity is clearly visible as oscillations 
   in the proper distance. }
 \end{figure}

\subsection{Initial data construction}
\label{sec:methods:ID}

\begin{figure}[t]
   \includegraphics[width=0.5\textwidth]{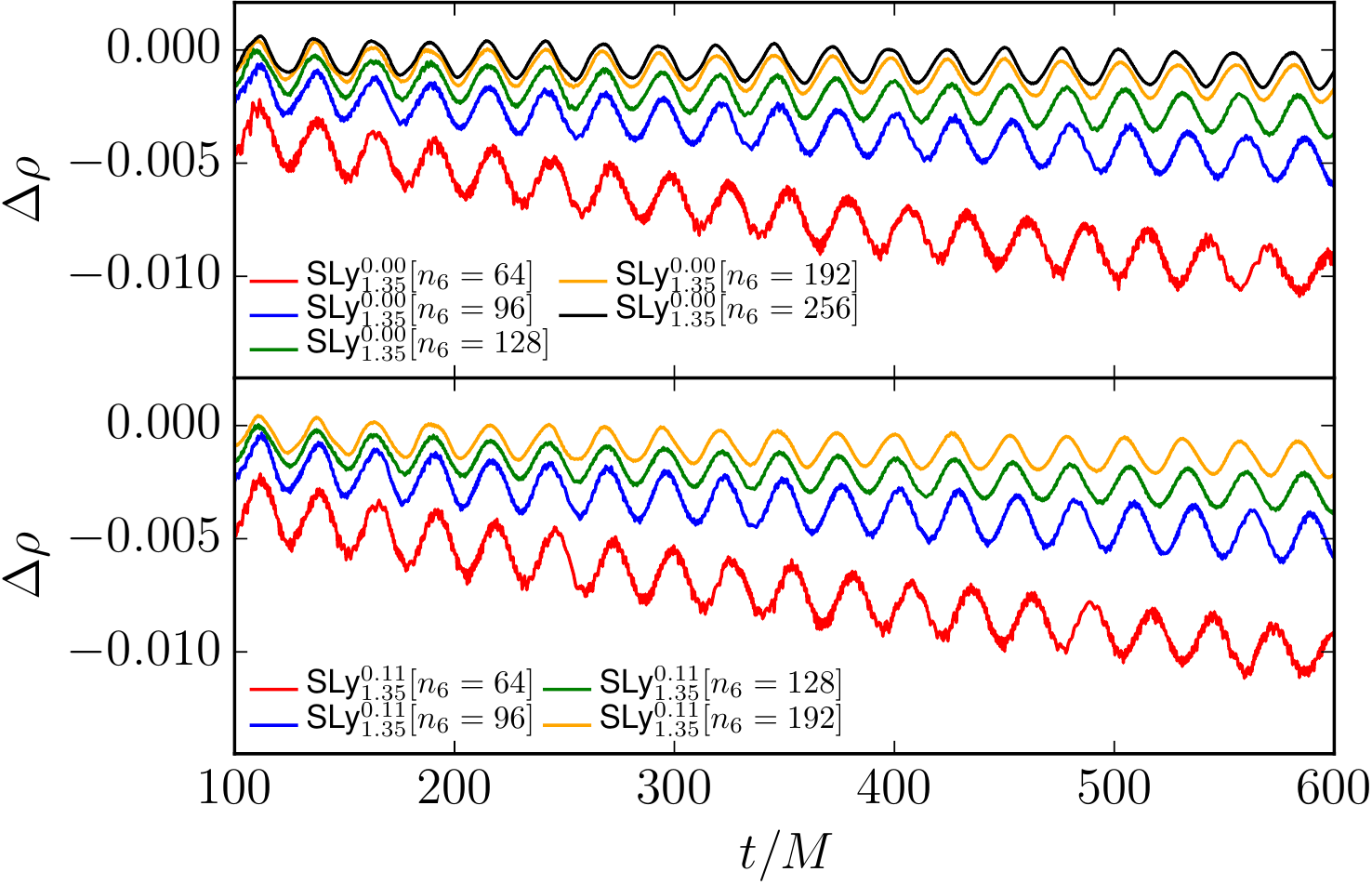}
   \caption{ \label{fig:density_oscillations}
   Central density oscillations for the non-spinning configuration 
   SLy$_{1.35}^{0.00}$ (top panel) 
   and the spinning configuration SLy$_{1.35}^{0.11}$ (bottom panel).
   The density oscillations are independent of the intrinsic spin of the 
   NSs which shows the validity of the constructed initial data 
   employing the CRV approach. }
 \end{figure}
 
For construction of the initial data, the pseudospectral 
SGRID code~\cite{Tichy:2006qn,Tichy:2009yr,Tichy:2009zr} is used. 
SGRID allows the computation of generic neutron star binary configurations
by combining the conformal thin sandwich 
equations~\cite{Wilson:1995uh,Wilson:1996ty,York:1998hy} together with
the constant rotational velocity (CRV)
approach~\cite{Tichy:2011gw,Tichy:2012rp,Tichy:2016vmv}.

The code uses pseudospectral methods to accurately 
compute spatial derivatives. The computational
domain is split into six patches including spatial infinity such that 
exact boundary conditions can be imposed. 
Due to the pseudospectral nature of the code, only relatively few grid
points are needed to reach the precision required for our simulations.
Thus SGRID does not need much memory, and we run it on a single compute node
or a workstation. The most computational expensive routines of SGRID are OpenMP
parallelized.

To obtain initial data with reduced eccentricity we employ 
an iterative procedure varying the binary's initial 
radial velocity and the eccentricity parameter until 
the eccentricity reaches a certain threshold, 
see~\cite{Dietrich:2015pxa}.
Figure~\ref{fig:ecc-red} presents one example of the eccentricity reduction 
procedure for the SLy$_{1.35}^{0.05}$ configuration. 
In most cases, three iteration steps are sufficient to 
obtain final eccentricities $\lesssim 10^{-3}$.\\

In addition to its capability to compute eccentricity reduced 
initial configurations, SGRID's advantage is that 
constraint solved initial data in hydrodynamical equilibrium can be computed 
for spinning NSs.
Solving the coupled system of elliptic equations is a challenging task, but 
required to achieve the necessary accuracy to allow gravitational waveform modeling. 
As an indicator for the accuracy of the numerical simulations 
we present central density ($\rho_c$) oscillations in Fig.~\ref{fig:density_oscillations}. 
In particular, we compute density oscillations
\begin{equation}
 \Delta \rho = \frac{\rho_c(t) - \rho_c(t=0)}{\rho_c(t=0)}. 
\end{equation} 
for SLy$_{1.35}^{0.00}$ (top panel) and SLy$_{1.35}^{0.11}$ (bottom panel).
The time evolution of $\Delta \rho$ is characterized by two 
effects (i) an almost linear decrease of the central density caused by the numerical 
discretization and (ii) oscillations caused by assumptions of the initial data. 
While effect (i) clearly decreases with increasing resolution, 
effect (ii) is almost independent of the resolution. 
The oscillations are of the order of $0.075\%$.
For systems not in hydrodynamical equilibrium those density oscillations can 
be about $20-30\%$, see e.g.~the study in~\cite{Moldenhauer:2014yaa,Dietrich:2015pxa}.
Important for our further consideration is that independent of 
the intrinsic rotation of the individual stars 
the magnitude of the oscillations remains unchanged, cf.~bottom panel of 
Fig.~\ref{fig:density_oscillations} in which we show the central density oscillations 
for SLy$_{1.35}^{0.11}$. 

\subsection{Dynamical Evolution}
\label{sec:methods:simulations}

\begin{figure}[t]
   \includegraphics[width=0.5\textwidth]{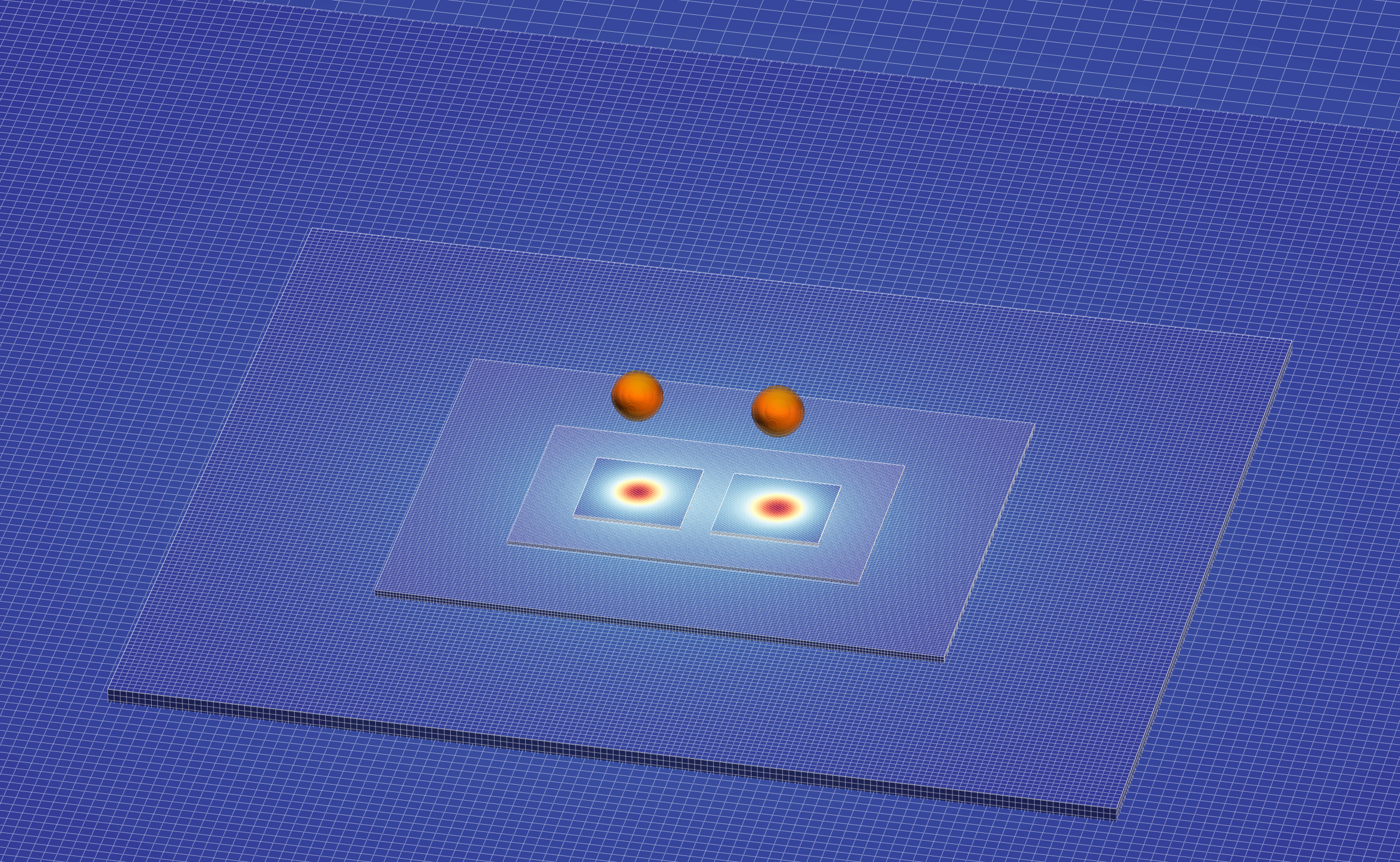}
   \caption{Example of the adaptive mesh refinement in BAM.
     We also show isocontour surfaces of the NS density and the metric's lapse function, that roughly indicates the gravitational potential. 
   \label{fig:mesh}}
 \end{figure} 

We employ the 
BAM code~\cite{Brugmann:2008zz,Thierfelder:2011yi,Dietrich:2015iva,Bernuzzi:2016pie}
for our simulations.
The code contains state-of-the-art methods to deal with general relativistic
hydrodynamics (GRHD) as well as black hole spacetimes. 
Finite difference stencils are used for the spatial 
discretization of the metric variables, 
and high resolution shock-capturing methods 
for the hydrodynamic variables.
The evolution algorithm is based on the method of lines.

The code uses an adaptive mesh refinement (AMR) technique in which  
the domain consists of a hierarchy of nested Cartesian grids (refinement levels). 
The grid spacing of each refinement level is half 
the grid spacing of its surrounding coarser refinement level.
The innermost levels move dynamically during the time evolution 
following the motion of the neutron stars such that 
the strong field region is always covered with the highest resolution.
We show a sketch of the refinement strategy in Fig.~\ref{fig:mesh}.
For the far field region, we have the option to use a ``cubed-sphere'' multi-patch AMR, 
which is particularly suited to accurately simulate the distant wave
zone. 
However, to save computational costs we do not employ the cubed-sphere multi-patch AMR
in this work. The grid configuration of the presented simulations consists of
a total of 7 refinement levels labeled $l=0,...,6$ with increasing resolution.
The specific grid setup is summarized in Table~\ref{tab:config}. 

The BAM code evolves spacetimes using either the 
BSSN~\cite{Nakamura:1987zz,Shibata:1995we,Baumgarte:1998te} or 
Z4c~\cite{Bernuzzi:2009ex,Hilditch:2012fp}
formulations. For the simulations proposed in this article, the 
Z4c evolution system is employed, which 
leads to an improved accuracy compared to the BSSN formulation, 
see e.g.~\cite{Ruiz:2010qj,Weyhausen:2011cg}. 

The numerical fluxes for the GRHD system are built using a
flux-splitting approach based on the local Lax-Friedrich flux
and performing the reconstruction on the characteristic
fields~\cite{Jiang:1996,Suresh:1997,Mignone:2010br}. 
For the reconstruction we use the fifth-order
WENOZ algorithm~\cite{Borges:2008a}. 
We refer to~\cite{Bernuzzi:2016pie} for further information about the 
GRHD implementation.\\

The code employs a hybrid OpenMP/MPI parallelization strategy.
Each refinement level is divided into equally sized
sub-boxes with ghost zones, whose sizes depend on the applied
finite-differencing stencil. 
Each of the sub-boxes is then owned and evolved
by a single MPI process. The ghost zones are synchronized after each
evolution step. Thus, each MPI process owns exactly one sub-box of
every mesh refinement box and works on the same number of grid points. 
This optimizes load balancing.
In addition, each MPI process launches an equal
number of OpenMP threads using shared memory. 
This helps to decrease the required memory, 
since fewer MPI processes with fewer ghost zones are needed.

\begin{figure}[t]
   \includegraphics[width=0.48\textwidth]{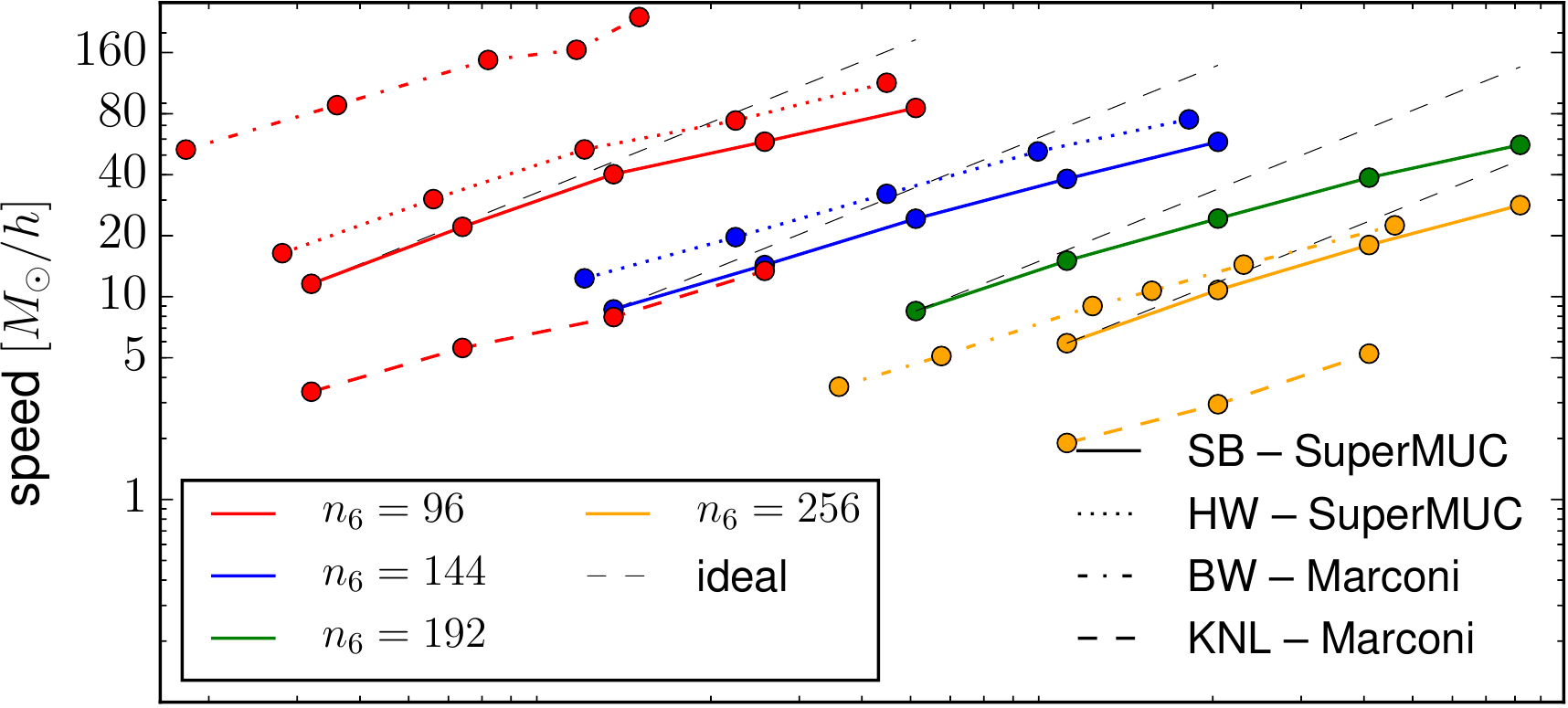}
   \includegraphics[width=0.48\textwidth]{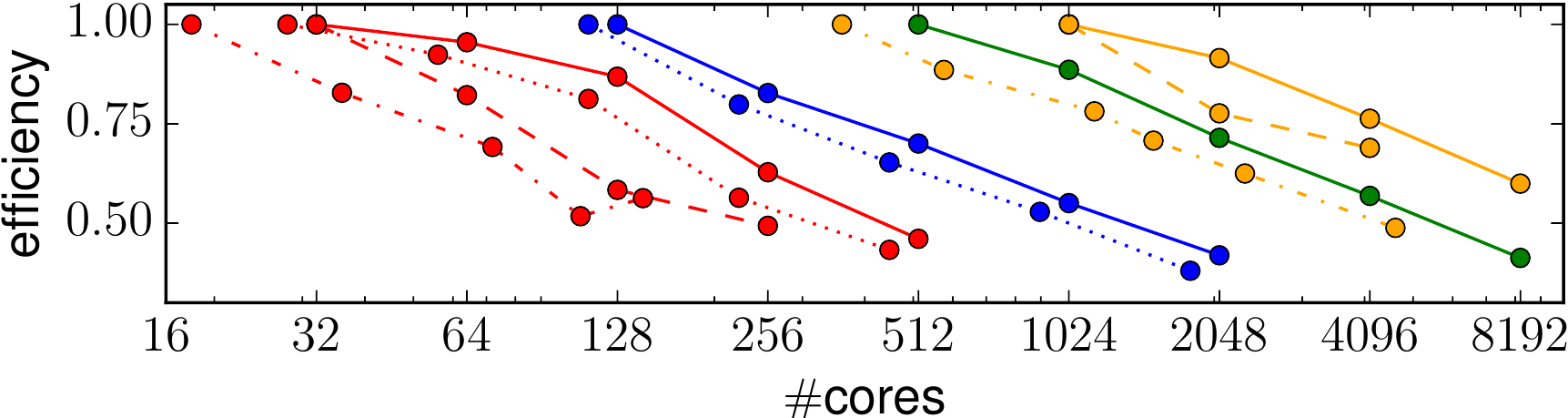}
   \caption{ \label{fig:scaling}
   Strong scaling (top) and efficiency (bottom) of the BAM code for different resolutions on Marconi and SuperMUC. 
   The test has been performed on the phase1 of SuperMUC with Sandy Bridge-EP
   Xeon E5-2680 8C processors (solid lines),   
   phase 2 of SuperMUC using Haswell Xeon Processor E5-2697v3 (dotted lines), on 
   the Intel Knights Landing partition of Marconi (dashed lines), 
   and on the Broadwell partition of Marconi (dot-dashed lines) with 
   Intel Xeon E5-2697 v4. 
   For reference ideal scaling is shown 
   for the simulations on phase 1 of SuperMUC.}
\end{figure} 

The evolution of the Einstein equations and relativistic hydrodynamics
uses approximately 150-200 grid variables
(including storage of previous timesteps) of double precision data type,
i.e.~8 bytes per grid point and variable.
Together with additional variables not directly used in the main evolution of the Einstein equations,
and taking into account (i) additional requirements for MPI buffer zones,
as well as
(ii) the fact that memory usage varies during evolution due the AMR
regridding, we estimate that the most demanding configurations need a
maximum of about $\sim 2.5$ TB.

\section{Parallel Performance \& Computational Resources}
\label{sec:performance}

Let us discuss the parallel performance of the BAM code in production
runs for BNS spacetime.

In the top panel of Fig.~\ref{fig:scaling} we present strong scaling 
results of the BAM code obtained on the
HPC system SuperMUC at the Leibniz Supercomputing Centre 
of the Bavarian Academy of Sciences and 
Humanities and on the Italian Tier-0 supercomputer Marconi at CINECA. 
We used the Intel 16 compiler on SuperMUC and on the Broadwell partition of Marconi 
and the Intel 17 compiler on Marconi's Knights Landing partition.
The strong scaling test is based on production-like simulations consisting of
two NSs covered by a total of seven refinement levels
with up to level $l=6$. We consider different grid setups and employ
$n^3=96^3,144^3,192^3,256^3$ points in each inner level, the outermost
levels use as usual twice the number of points in each direction for
up to $512^3$ points. 

The plot shows speed (top panel) and efficiency
(bottom panel) defined as 
\begin{equation}
{\rm efficiency} =\frac{{\rm speed}_2/{\rm speed_1}}{\#{\rm cores}_2/\#{\rm cores}_1}.
\end{equation}
The efficiency in the bottom panel of Fig.~\ref{fig:scaling} refers to 
the smallest number of cores on which the grid configuration 
has been tested on.

We find that for small jobs the code speed is best on the Broadwell 
partition of Marconi, although efficiency is better on SuperMUC's Haswell 
and Sandy Bridge nodes.
The largest runs have comparable performances on Broadwell and Sandy Bridge.
Efficiency drops below 75\% between 2048 and 4096 cores depending on 
the machine.
Interpretation of the scaling results on Knights Landing architectures
requires some care. Such processors have approximately half clock
speed than the others and peak performance can be only reached
exploiting large vector instructions and optimizing memory loads. Although they anticipate some of
the features required for Exascale Computing (e.g. power efficiency
with a large FLOP-s/watt ratios), significant code refactoring is
needed for an optimal usage.
The BAM code, in particular, is optimized for more traditional
architectures and main routines are not vectorized. In our experience the typical
speed is lower by a factor about 5-7 in most of the applications; the
performances are significantly worse for small jobs. \\

The simulations presented here are the largest BNS simulations 
performed with the BAM code so far. 
We used allocations at the HPC clusters 
SuperMUC (Germany), 
JURECA (Germany), 
Stampede (US), 
and Marconi (Italy), and utilized a total $\sim 25$ million CPU hours. 
Our largest production runs use $n=256$ but an optimal setup for simulations with $n=320$
is currently under investigation with preliminary results ongoing.
The most demanding setups run on about 1000 compute cores with a total runtime of 
approximately 2-3 months. 

\section{Waveform Error budget}
\label{sec:accuracy}

\begin{figure}[t]
   \includegraphics[width=0.48\textwidth]{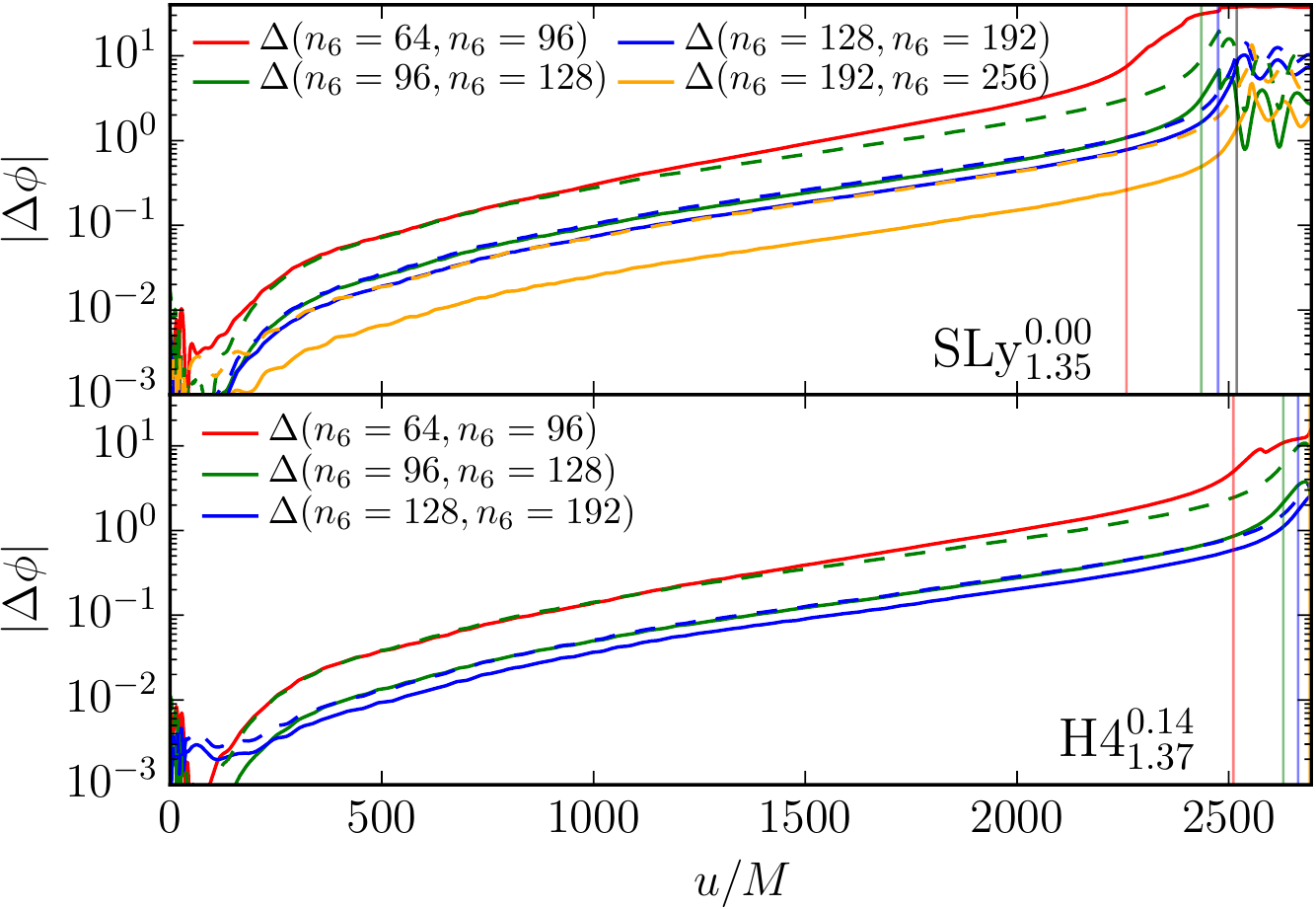}
   \caption{ \label{fig:convergence_phase}
   Phase differences between different resolutions versus retarded time
   for configurations
   SLy$_{1.35}^{0.00}$ (top panel) and H4$_{1.37}^{0.14}$ (bottom panel).
   The rescaling factor is chosen such that each rescaled (i.e. dashed) line
   would coincide with the line from the next lower resolution phase
   difference for exact second order convergence.
   The vertical lines mark the moment
   of merger (peak of the GW amplitude) for different resolutions. }
 \end{figure} 

The main uncertainties in GWs obtained from numerical simulations are 
(I) errors caused by the discretization of the underlying continuum problem and 
(II) errors due to the extraction of GWs at finite radii. 
An additional source of error is the artificial atmosphere (necessary
for the simulation of low density regions) and the related possible
violation of rest mass conservation. Fortunately, this effect is 
subdominant and 
converges to zero with increasing resolution 
and it is therefore included in (I). 

Considering error (I) any finite-differencing approximation $f^{(h)}$
computed at resolution $h$ can be written as 
\begin{equation}
f^{(h)} = f^{(e)} + \sum^{\infty}_{i=p} A_i h^i \ ,
\end{equation}
where $f^{(e)}$ is the continuum solution for
$h\to0$, and $p$ the convergence order. 
Although it is impossible to compute $f^{(e)}$ one can extrapolate from 
a dataset with different resolutions to obtained a more accurate result.
A way to proceed is to consider Richardson extrapolation~\cite{Richardson:1911}. This requires a dataset $(f^{(h)})$ at different finite resolutions and 
an accurate measure of the convergence order $p$.

While BAM's flux scheme based on reconstruction of characteristic
fields~\cite{Jiang:1996,Suresh:1997,Mignone:2010br} and 
a WENOZ scheme~\cite{Borges:2008a} can achieve 
fifth order convergence for smooth matter fields, 
we do not find such high convergence orders for our simulations. 
Ref.~\cite{Bernuzzi:2016pie} pointed out that the surface of the neutron star, 
which is only $C^0$ continuous, leads to non-ideal weights in the 
WENOZ reconstruction and consequently to second order convergence. 

Second order convergence is, however, achieved independently of the particular details of the 
numerical simulation, i.e.~configuration parameters and grid setup. 
As exemplary cases we present phase differences for SLy$_{1.35}^{0.00}$ (top panel) 
and H4$_{1.37}^{0.14}$ (bottom panel) in Fig.~\ref{fig:convergence_phase}.
No alignment of the waveforms has been performed at the beginning of the
simulations. For both setups second order convergence is achieved almost 
up to the moment of
merger. We mark the moment of merger (the time where the GW amplitude peaks)
for different resolutions by vertical colored lines.
This allows us to use our datasets for Richardson extrapolation and construct a 
more accurate waveform.\\

To mitigate errors due to extraction at a finite radius (II),
we evaluate the waveform at different radii $r_j$ with $j=0...N$ and
extrapolate the phase and amplitude to infinite radius
using a polynomial of order $K<N$, 
\begin{equation}
\label{poly_extrar}
f(u; r_j) = f_0(u) + \sum^K_{k=1} f_k(u) r^{-k}_j \ .
\end{equation}

\begin{figure}[t]
   \includegraphics[width=0.48\textwidth]{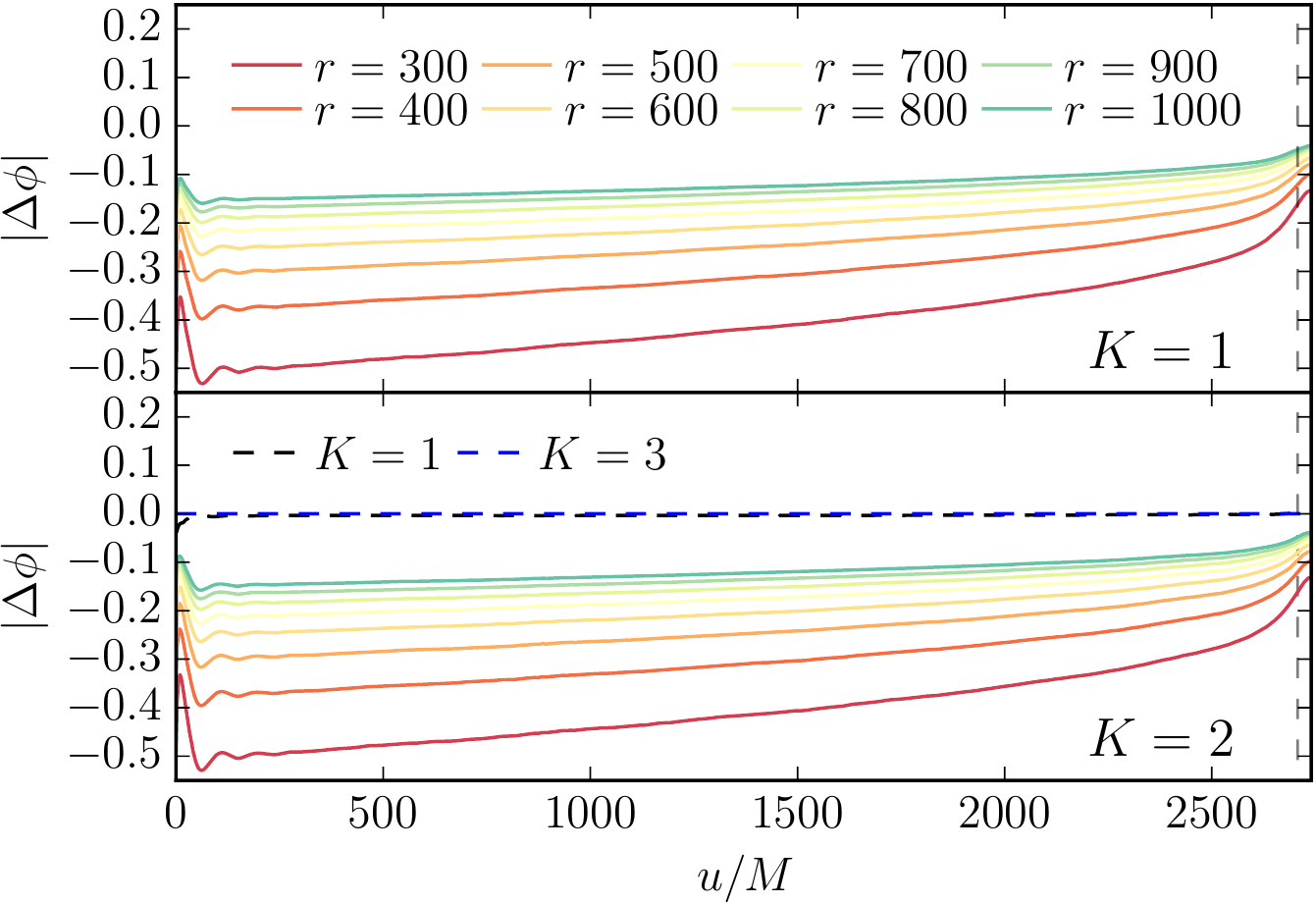}
   \caption{
   Phase difference between waveforms extracted at various finite radii
   and extrapolated waveforms assuming 
   $K=1$ (top panel) and $K=2$ (bottom panel). 
   In the bottom panel also the phase difference between different 
   extrapolation orders is shown.
   \label{fig:extrapolation}}
 \end{figure}

Figure~\ref{fig:extrapolation} present as an exemplary case the phase differences between 
finite radii extracted waveforms and polynomial extrapolated waveforms with $K=1$ (top panel) and $K=2$ (bottom panel) for MS1b$_{1.35}^{0.10}$.
We obtain similar results independent of the chosen extrapolation order, cf.~bottom panel of 
Fig.~\ref{fig:extrapolation} for the phase differences between extrapolation order $K=1,K=2$, and $K=3$. 
However, for higher extrapolation orders we find spurious oscillations
and that the extrapolation is less robust, i.e.~it depends more on the
chosen extraction radii. Therefore, we have decided to use $K=2$ extrapolation. 
The error is estimated by computing the 
phase difference with respect to a waveform extracted at finite radius
of $r=1000$. Notice that this error is a conservative estimate 
since also larger extraction radii are available to validate the results and 
the phase difference between different extrapolation orders is basically zero. 
While during the inspiral the finite radii extrapolation dominates the overall 
error budget, close to the merger errors due to the numerical discretization dominate. 

\section{Applications}
\label{sec:applications}

\begin{figure}[t]
   \includegraphics[width=0.48\textwidth]{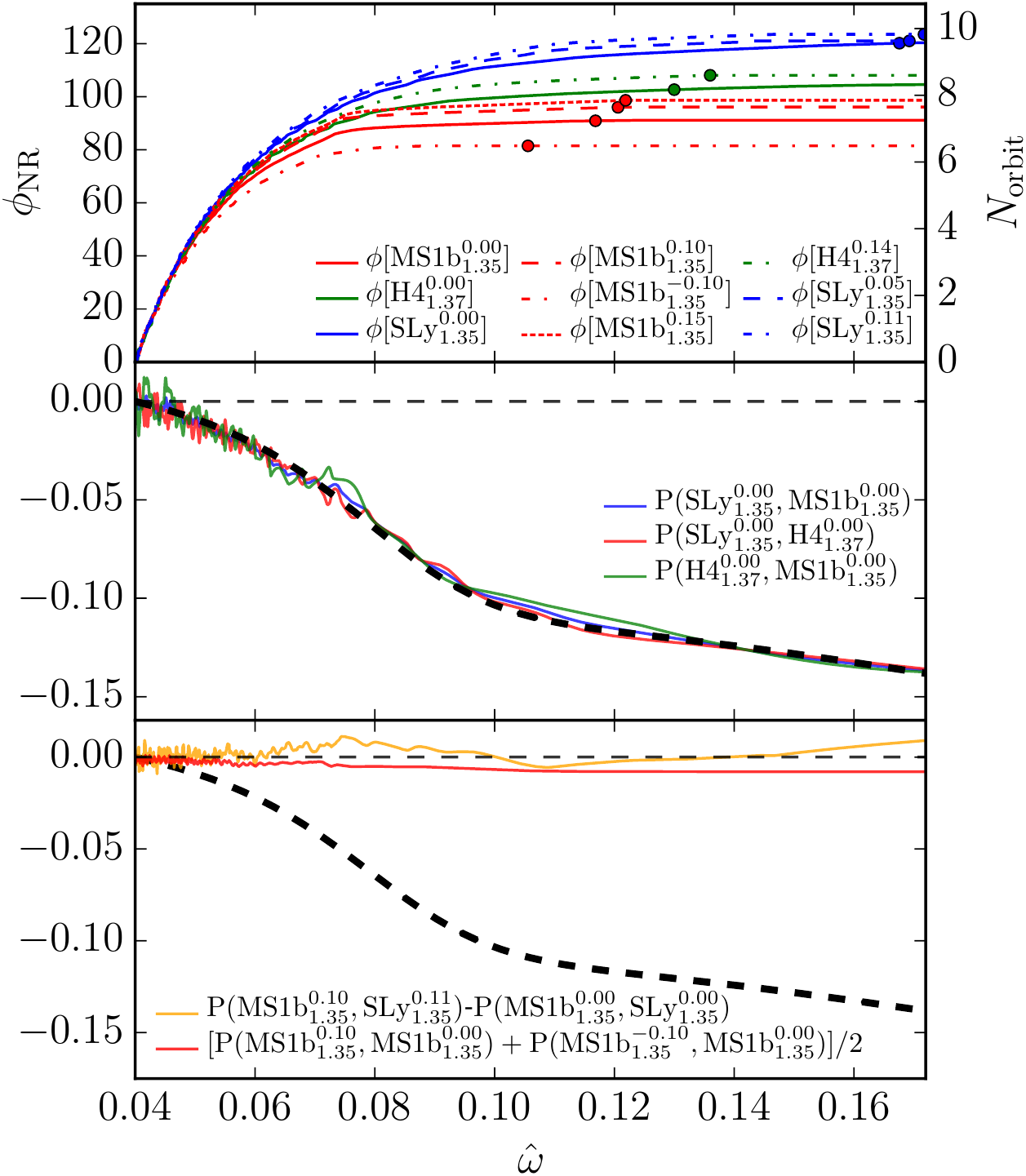}
   \caption{Top panel: 
   Accumulated phase/number of orbits of the numerical relativity simulations with respect to
   a dimensionless frequency of $\hat{\omega}=0.04$, which corresponds to about 
   2 orbits after the start of the simulations. 
   Middle panel: Estimates for $f(\hat{\omega})$, Eq.~\eqref{eq:phiT}, 
   for combinations of different simulations. 
   The dashed black line is the estimate of $f(\hat{\omega})$ used 
   in the \NRtidal{} model.
   Bottom panel: Combinations of different simulations to extract 
   the spin-spin contribution to the phase in the late inspiral (red)
   as well as to estimate possible tidal-spin couplings (orange). 
   For comparison, we include again the estimate of $f(\hat{\omega})$ of the \NRtidal{} model as a black dashed line. 
   \label{fig:phiomega}}
 \end{figure} 

\subsection{GW templates construction}

GW data analysis pipelines generate
a large number, $\sim10^6-10^7$, of waveform templates that are then
cross-correlated with the signal. As a consequence, waveform models
that can be produced in a fast and efficient way must be in place.
In Ref.~\cite{Dietrich:2017aum} we presented the first closed form 
tidal approximant combining Post-Newtonian (PN), 
effective-one-body (EOB), and numerical relativity results.
[We call the approximant in the following \NRtidal{} 
which is the name used in the LSC Algorithm Library Suite.] 

The model can be added to any binary black hole
baseline and describes the GW phase evolution 
due to tidal effects during the inspiral up to merger.
The two main assumptions for the model are:
(i) tidal effects are proportional to the effective tidal coupling 
constant~\cite{Damour:2009wj,Damour:2012yf} 
\begin{equation}\label{eq:kappa}
\kappa^T_{\rm eff} = \frac{2}{13} \left[ 
\left(1+12\frac{X_B}{X_A}\right)\left(\frac{X_A}{C_A}\right)^5 k^A_2 + 
 (A \leftrightarrow B) 
\right]  \ ,
\end{equation}
where $k^A_2$ is the quadrupolar Love number describing the static
quadrupolar deformation of one body in the gravitoelectric field of
the companion, $X_A=M_A/M$, and $C_A$ is the compactness of star $A$, 
and (ii) tidal and spin effects can be decoupled. \\

We now check the validity of our assumptions on the simulation data.
In the top panel of Fig.~\ref{fig:phiomega} 
we present the dimensionless GW phase as a function 
of the GW frequency. Additionally, we extract 
the tidal contribution to the phase 
\begin{equation}
\phi_T = \kappa_{\rm eff}^T \cdot f(\hat{\omega})  \label{eq:phiT}
\end{equation}
with $\hat{\omega}$ being the dimensionless GW frequency
from 
\begin{equation}
 {\rm P}({\rm BNS}_A, {\rm BNS}_B) = 
 \frac{\phi({\rm BNS}_A)-\phi({\rm BNS}_B)}
 {\kappa_{\rm eff}^T({\rm BNS}_A) - \kappa_{\rm eff}^T({\rm BNS}_B) }.
 \label{eq:P}
\end{equation}

The middle panel of Fig.~\ref{fig:phiomega}
shows ${\rm P}$ for all combinations of irrotational 
NS configurations. This together with Eq.~\eqref{eq:phiT} 
allows us to present an estimate for $f(\hat{\omega})$, which 
is the most important quantity in the \NRtidal{} model 
(dashed black line in the middle and bottom panel of Fig.~\ref{fig:phiomega}). 
According to Fig.~\ref{fig:phiomega} all combinations 
lead to a similar result for $f(\hat{\omega})$
which validates assumption~(i). \\

The bottom panel of Fig.~\ref{fig:phiomega} shows
combinations of simulations including spinning configurations. 
Combining spin aligned, anti-aligned, and non-spinning 
data of the same EOS allows us to estimate spin-spin interactions 
(red line). Clearly visible is that the spin-spin interactions are 
almost zero and not well resolved in our simulations. 
This is important since information about the EOS 
are encoded in spin-spin interaction describing the deformation of the NS 
due to its intrinsic rotation.  
Additionally, we show a combination computed with approximately
the same spin magnitudes but different EOSs (orange curve). 
If appreciable spin-tidal coupling was present in our simulations, 
the curve would deviate from zero. Since both red and orange curves
are so close to zero, both spin-tide coupling and the imprint of 
the EOS on the spin-spin interaction are small.  
Future simulations with even larger resolutions or possibly higher 
spin magnitudes might reveal these effects, but current state-of-the art 
simulations suggest that a decoupling of spin and tidal effects 
[assumption (ii)] is 
valid. This significantly simplifies waveform development. 

Knowing the time domain behavior of $f(\hat{\omega})$
the frequency-domain phase correction in the \NRtidal{} is constructed as follows, 
first, fitting $f(\hat{\omega})$ with a Pad\'e approximant (black dashed line in Fig.~\ref{fig:phiomega}), 
second, computing the frequency domain tidal phase numerically under the 
stationary phase approximation~\cite{Damour:2012yf},
and, finally, fitting the numerical result again with a Pad\'e approximant. 
The resulting \NRtidal{} model thus gives both a closed-form 
expression for the time domain and frequency domain tidal corrections.

\subsection{GW template verification}

\begin{figure}[t]
   \includegraphics[width=0.48\textwidth]{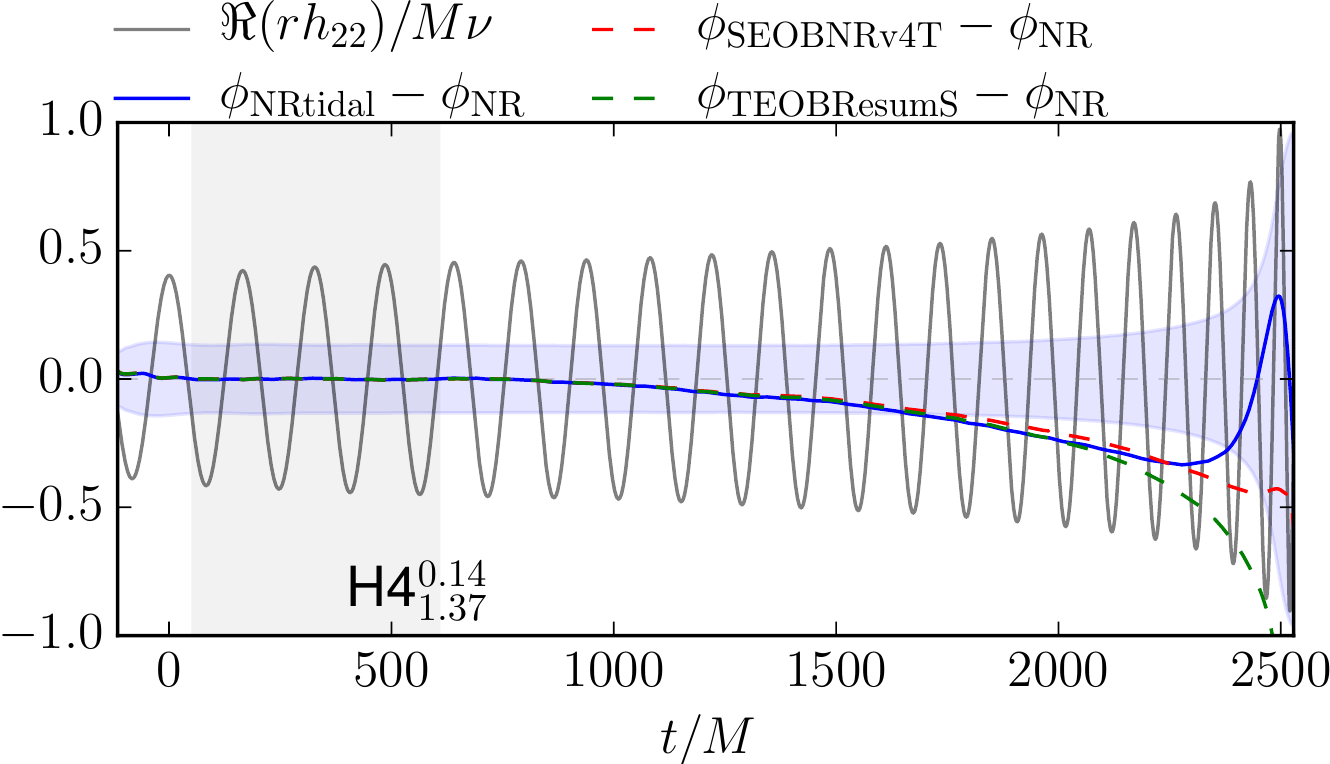}
   \caption{Comparison of the H4$_{\rm 1.37}^{0.14}$ configuration 
   with the \NRtidal{} (blue), \texttt{SEOBNRv4T} (red), \texttt{TEOBResumS} (green) waveform models.
   Numerical uncertainty is shown as a blue shaded region. 
   As a gray curve we include the real part of the numerical waveform. 
   The waveform approximants are aligned to the numerical relativity 
   waveform in the gray shaded region. 
   \label{fig:comparison}}
 \end{figure} 

A second usage of the simulation data is the verification of
different waveform models. In the following we compare our data to 
the previously introduced \NRtidal{} approximant 
as well as two distinct EOB models with spin aligned to the angular momentum and tidal 
effects \texttt{SEOBNRv4T}~\cite{Hinderer:2016eia,Steinhoff:2016rfi}
and \texttt{TEOBResumS}~\cite{Bernuzzi:2014owa,Damour:2014sva,Nagar:2015xqa,NagarinPrep}.

We show results for the particular H4$_{\rm 1.37}^{0.14}$ configuration 
in Fig.~\ref{fig:comparison}. 
We find that all the approximants deviate significantly from NR in the representation of 
last few orbits ($t/M\gtrsim2000$). However, the deviation is small: $\sim0.1$~rad for 
\NRtidal{} and \texttt{SEOBNRv4T} approximant and $\sim0.4$~rad for \texttt{TEOBResumS}. 
Note the NR error is about $\pm1{\rm rad}$ up to the moment of merger. Our comparison shows that 
with state-of-the-art techniques, numerical relativity reaches an accuracy 
at which calibration of tidal EOB models to simulations becomes possible.

\section{Conclusion}
\label{sec:conclusion}

In this article we presented new high resolution 
simulations of binary neutron stars.  
%Due to the significant progress in numerical relativity 
We are able to simulate spinning neutron star binaries and produce waveform that convergence in rigorous self-convergence tests. Using extrapolation based on convergence tests we compute waveforms with phase uncertainties of $\sim1$~rad, accumulated over $\sim12$ orbits. 
The simulations have been performed on a variety of 
HPC systems and required about $25$ million CPU hours. 
Higher resolution runs are ongoing and will allow 
us to reduce numerical uncertainties even further. 

With regard to numerical methods, we are exploring the feasibility of
a next generation code,
bamps~\cite{Bugner:2015gqa,Hilditch:2015aba,Ruter:2017iph}, that
implements discontinuous Galerkin (DG) methods for numerical
relativity and general relativistic hydrodynamics. In principle, such
methods allow high-order schemes combined with parallel AMR, with
scaling to a much larger number of processors than reported here,
although a full-featured implementation of DG methods for binary
neutron star simulations does not exist yet.

With the numerical methods employed in BAM, second order 
convergence is achieved for the gravitational 
wave phase for all configurations. The absence of higher order 
convergence is caused by the presence of discontinuities at the 
neutron star surface. However, the clean second order convergence
allows a proper error estimate and the use of Richardson 
extrapolation to obtain improved representations of the 
continuum solution. Overall, accurate numerical relativity waveforms  are urgently needed to 
further develop, improve, and test waveform 
models and to be prepared for future gravitational wave detections 
in the new era of gravitational wave astronomy.

%%______________________________________________________________

\section*{Acknowledgments}
  T.D. acknowledges support by the European Union’s Horizon 
  2020 research and innovation program under grant
  agreement No 749145, BNSmergers.
  S.B.~acknowledges support by the European Union's H2020 
  under ERC Starting Grant, grant
  agreement no. BinGraSp-714626. 
  W.T.~was supported by the National Science Foundation under grants
  PHY-1305387 and PHY-1707227.
  Computations were performed on SuperMUC at the LRZ (Munich) under 
  the project number pr48pu, JURECA (J\"ulich) 
  under the project number HPO21, Stampede 
  (Texas, XSEDE allocation - TG-PHY140019), 
  Marconi (CINECA) under ISCRA-B the project number HP10BMAB71, and
  under PRACE allocation from the 14th Tier-0 call.

\bibliographystyle{IEEEtran}
\bibliography{paper20180321.bbl}

\end{document}